\documentclass[aps,prl,reprint,groupedaddress]{revtex4-1}
\usepackage{graphicx} 
\usepackage{dcolumn} 
\usepackage{bm}
\usepackage{hyperref}

\begin{document}

\title{Constraining the high-density behavior of nuclear equation of state from strangeness production in heavy-ion collisions}

\author{Zhao-Qing Feng}
\email{fengzhq@impcas.ac.cn}
\affiliation{Institute of Modern Physics, Chinese Academy of Sciences, Lanzhou 730000, People's Republic of China}

\date{\today}

\begin{abstract}
The dynamics of pions and strange particles in heavy-ion collisions in the region of 1A GeV energies is investigated by the lanzhou quantum molecular dynamics model for probing the nuclear equation of state at supra-saturation densities. The total multiplicities and the ratios obtained in $^{197}$Au+$^{197}$Au over $^{12}$C+$^{12}$C systems are calculated for selected Skyrme parameters SkP, SLy6, Ska and SIII, which correspond to different modulus of incompressibility of symmetric nuclear matter and different cases of the stiffness of symmetry energy. A decrease trend of the excitation functions of the ratios for strange particle production with increasing incident energy was observed. The available data of K$^{+}$ production measured by KaoS collaboration are described well with the parameter SkP, which results in a soft equation of state. The conclusions can not be modified by an in-medium kaon-nucleon potential.
\begin{description}
\item[PACS number(s)]
25.75.-q, 13.75.Gx, 25.80.Ls
\end{description}
\end{abstract}

\maketitle

Particle production in relativistic heavy-ion collisions has been investigated as a useful tool to extract the information of the nuclear equation of state (EoS) under extreme conditions in terrestrial laboratories, such as high density, high temperature and large isospin asymmetry etc. Besides nucleonic observables such as rapidity distribution and flow of free nucleons or light clusters (such as deuteron, triton and alpha etc.), also mesons emitted from
the reaction zone can be probes of the hot and dense nuclear matter. In particular, the strangeness productions (K$^{0,+}$, $\Lambda$ and $\Sigma^{-,0,+}$) in heavy-ion collisions in the region of 1A GeV energies have been extensively investigated both experimentally \cite{Fo07,St01,Me07} and theoretically \cite{Ai85,Li97,Fu01,Ha06}. Kaons ($K^{0}$ and $K^{+}$) as a probe of EoS are produced in the high density phase without subsequent reabsorption process. It was noticed that the kaon yields are sensitive to the EoS in theoretical investigations by transport models. The available data already favored a soft EoS at high densities. The $K^{0}/K^{+}$ ratio was also proposed as a sensitive probe to constrain the high-density behavior of the symmetry energy \cite{Fe06,Pr10}. It is not only in understanding the reaction dynamics, the high-density behavior of the EoS also has an important application in astrophysics, such as the structure of neutron star, the cooling of protoneutron stars, the nucleosynthesis during supernova explosion of massive stars etc \cite{St05,Kl06}.

The lanzhou quantum molecular dynamics (LQMD) model has been successfully applied to treat the dynamics in heavy-ion fusion reactions near Coulomb barrier and also to describe the capture of two heavy colliding nuclides to form a superheavy nucleus \cite{Fe05,Fe08}. Further improvements of the LQMD model have been performed in order to investigate the dynamics of pion and strangeness productions in heavy-ion collisions and to extract the information of isospin asymmetric EoS at supra-saturation densities \cite{Fe09,Fe10a,Fe10b,Fe10c}. In this work, the high-density behavior of nuclear equation of state is to be investigated through particle productions in heavy-ion collisions in the region of 1A GeV energies.

In the LQMD model, the time evolutions of the baryons and mesons in
the system under the self-consistently generated mean-field are
governed by Hamilton's equations of motion, which read as
\begin{eqnarray}
\dot{\mathbf{p}}_{i}=-\frac{\partial H}{\partial\mathbf{r}_{i}},
\quad \dot{\mathbf{r}}_{i}=\frac{\partial
H}{\partial\mathbf{p}_{i}}.
\end{eqnarray}
Here we omit the shell correction part in the Hamiltonian $H$ as
described in Ref. \cite{Fe08}. The Hamiltonian of baryons consists
of the relativistic energy, the effective interaction potential energy and
the momentum dependent part as follows:
\begin{equation}
H_{B}=\sum_{i}\sqrt{\textbf{p}_{i}^{2}+m_{i}^{2}}+U_{int}+U_{mom}.
\end{equation}
Here the $\textbf{p}_{i}$ and $m_{i}$ represent the momentum and the
mass of the baryons. The momentum dependent term is taken as the same form in Ref. \cite{Ai87}, which reduces the effective mass in nuclear medium.

The effective interaction potential is composed of the Coulomb
interaction and the local interaction
\begin{equation}
U_{int}=U_{Coul}+U_{loc}.
\end{equation}
The Coulomb interaction potential is calculated by
\begin{equation}
U_{Coul}=\frac{1}{2}\sum_{i,j,j\neq
i}\frac{e_{i}e_{j}}{r_{ij}}erf(r_{ij}/\sqrt{4L})
\end{equation}
where $e$ is the charged number including protons and
charged resonances. The $r_{ij}=|\mathbf{r}_{i}-\mathbf{r}_{j}|$ is
the relative distance of two charged particles.

The local interaction potential energy is derived directly from the Skyrme
energy-density functional and expressed as
\begin{equation}
U_{loc}=\int V_{loc}(\rho(\mathbf{r}))d\mathbf{r}.
\end{equation}
The local potential energy-density functional reads
\begin{eqnarray}
V_{loc}(\rho)=&& \frac{\alpha}{2}\frac{\rho^{2}}{\rho_{0}}+
\frac{\beta}{1+\gamma}\frac{\rho^{1+\gamma}}{\rho_{0}^{\gamma}}+
\frac{g_{sur}}{2\rho_{0}}(\nabla\rho)^{2}  \nonumber \\
&& + \frac{g_{sur}^{iso}}{2\rho_{0}}[\nabla(\rho_{n}-\rho_{p})]^{2}  \nonumber \\
&& + \left(a_{sym}\frac{\rho^{2}}{\rho_{0}}+b_{sym}\frac{\rho^{1+\gamma}}{\rho_{0}^{\gamma}}+
c_{sym}\frac{\rho^{8/3}}{\rho_{0}^{5/3}}\right)\delta^{2}  \nonumber \\
&& + g_{\tau}\rho^{8/3}/\rho_{0}^{5/3},
\end{eqnarray}
where the $\rho_{n}$, $\rho_{p}$ and $\rho=\rho_{n}+\rho_{p}$ are
the neutron, proton and total densities, respectively, and the
$\delta=(\rho_{n}-\rho_{p})/(\rho_{n}+\rho_{p})$ is the isospin
asymmetry. Here, all the terms in the Skyrme energy functional are included in the model besides the spin-orbit coupling. The coefficients $\alpha$, $\beta$, $\gamma$, $g_{sur}$,
$g_{sur}^{iso}$, $g_{\tau}$ are related to the Skyrme parameters
$t_{0}, t_{1}, t_{2}, t_{3}$ and $x_{0}, x_{1}, x_{2}, x_{3}$ as \cite{Fe08},
\begin{eqnarray}
&& \frac{\alpha}{2}=\frac{3}{8}t_{0}\rho_{0}, \quad
\frac{\beta}{1+\gamma}=\frac{t_{3}}{16}\rho_{0}^{\gamma}, \\
&& \frac{g_{sur}}{2}=\frac{1}{64}(9t_{1}-5t_{2}-4x_{2}t_{2})\rho_{0}, \\
&& \frac{g_{sur}^{iso}}{2}=-\frac{1}{64}[3t_{1}(2x_{1}+1)+t_{2}(2x_{2}+1)]\rho_{0}, \\
&& g_{\tau}=\frac{3}{80}\left(\frac{3}{2}\pi^{2}\right)^{2/3}(3t_{1}+5t_{2}+4x_{2}t_{2})\rho_{0}^{5/3}.
\end{eqnarray}
The parameters of the potential part in the bulk symmetry
energy term are also derived directly from Skyrme energy-density
parameters as
\begin{eqnarray}
&& a_{sym}=-\frac{1}{8}(2x_{0}+1)t_{0}\rho_{0}, \quad b_{sym}=-\frac{1}{48}(2x_{3}+1)t_{3}\rho_{0}^{\gamma}, \nonumber \\
&& c_{sym}=-\frac{1}{24}\left(\frac{3}{2}\pi^{2}\right)^{2/3}\rho_{0}^{5/3}[3t_{1}x_{1}-t_{2}(5x_{2}+4)].
\end{eqnarray}
In the calculations, we use the Skyrme parameters SkP, SLy6, Ska and SIII, which result in different modulus of the incompressibility of symmetric nuclear matter and also different stiffness of the symmetry energy. In Table 1 we list the LQMD parameters related to several typical Skyrme forces.

\begin{table*}
\caption{\label{tab:table1} LQMD parameters and properties of symmetric nuclear
matter for Skyrme effective interactions after the inclusion of the
momentum dependent interaction.}
\begin{ruledtabular}
\begin{tabular}{ccccccccc}
&Parameters                 &SkM*   &Ska    &SIII   &SVI    &SkP    &RATP   &SLy6   \\
\hline
&$\alpha$ (MeV)             &-325.1 &-179.3 &-128.1 &-123.0 &-357.7 &-250.3 &-296.7 \\
&$\beta$  (MeV)             &238.3  &71.9   &42.2   &51.6   &286.3  &149.6  &199.3  \\
&$\gamma$                   &1.14   &1.35   &2.14   &2.14   &1.15   &1.19   &1.14   \\
&$g_{sur}$(MeV fm$^{2}$)    &21.8   &26.5   &18.3   &14.1   &19.5   &25.6   &22.9   \\
&$g_{sur}^{iso}$(MeV fm$^{2}$)&-5.5 &-7.9   &-4.9   &-3.0   &-11.3  &0.0    &-2.7   \\
&$g_{\tau}$ (MeV)           &5.9    &13.9   &6.4    &1.1    &0.0    &11.0   &9.9    \\
&$E_{sym}$ (MeV)            &30.1   &33.0   &28.2   &27.0   &30.9   &29.3   &32.0   \\
&$a_{sym}$ (MeV)            &62.4   &29.8   &38.9   &42.9   &94.0   &79.3   &130.6  \\
&$b_{sym}$ (MeV)            &-38.3  &-5.9   &-18.4  &-22.0  &-63.5  &-58.2  &-123.7 \\
&$c_{sym}$ (MeV)            &-6.4   &-3.0   &-3.8   &-5.5   &-13.0  &-4.1   &12.8   \\
&$\rho_{\infty}$ (fm$^{-3}$)&0.16   &0.155  &0.145  &0.144  &0.162  &0.16   &0.16   \\
&$K_{\infty}$ (MeV)         &215    &262    &353    &366    &200    &239    &230    \\
\end{tabular}
\end{ruledtabular}
\end{table*}

Analogously to baryons, the evolution of mesons (here mainly pions and kaons) is also determined by the Hamiltonian, which is given by
\begin{eqnarray}
H_{M}&& = \sum_{i=1}^{N_{M}}\left( V_{i}^{\textrm{Coul}} + \omega(\textbf{p}_{i},\rho_{i}) \right).
\end{eqnarray}
Here the Coulomb interaction is given by
\begin{equation}
V_{i}^{\textrm{Coul}}=\sum_{j=1}^{N_{B}}\frac{e_{i}e_{j}}{r_{ij}},
\end{equation}
where the $N_{M}$ and $N_{B}$ are the total numbers of mesons and
baryons including charged resonances. Here, we use the energy in vacuum for pions, namely $\omega(\textbf{p}_{i},\rho_{i})=\sqrt{\textbf{p}_{i}^{2}+m_{M}^{2}}$ with the momentum $\textbf{p}_{i}$ and the mass $m_{M}$ of the pions. We consider two scenarios for kaon (antikaon) propagation in nuclear medium, one with and one without medium modification. From the chiral Lagrangian the kaon and antikaon energy in the nuclear medium can be written as \cite{Li97,Ka86}
\begin{equation}
\omega(\textbf{p}_{i},\rho_{i})=\left[m_{K}^{2}+\textbf{p}_{i}^{2}-a_{K}\rho_{i}^{S}+
(b_{K}\rho_{i})^{2}\right]^{1/2}+b_{K}\rho_{i},
\end{equation}
\begin{equation}
\omega(\textbf{p}_{i},\rho_{i})=\left[m_{\overline{K}}^{2}+\textbf{p}_{i}^{2}-a_{\overline{K}}\rho_{i}^{S}+
(b_{K}\rho_{i})^{2}\right]^{1/2}-b_{K}\rho_{i},
\end{equation}
respectively. Here the $b_{K}=3/(8f_{\pi}^{2})\approx$0.32 GeVfm$^{3}$, the $a_{K}$ and $a_{\overline{K}}$ are 0.18 GeV$^{2}$fm$^{3}$ and 0.3 GeV$^{2}$fm$^{3}$, respectively, which result in the strength of repulsive kaon-nucleon potential and of attractive antikaon-nucleon potential with the values of 25.5 MeV and -96.8 MeV at saturation baryon density.

A hard core scattering in two particle collisions is assumed in the simulation of the collision processes by Monte Carlo procedures, in which the scattering of two particles is determined by a geometrical minimum distance criterion $d\leq\sqrt{0.1\sigma_{tot}/\pi}$ fm weighted by the Pauli blocking of the final states \cite{Ai91,Be88}. Here, the total cross section $\sigma_{tot}$ in mb is the sum of the elastic and all inelastic cross section. The probability reaching a channel in a collision is calculated by its contribution of the channel cross section to the total cross section as $P_{ch}=\sigma_{ch}/\sigma_{tot}$. The choice of the channel is done randomly by the weight of the probability. The primary products in nucleon-nucleon (NN) collisions at the 1A GeV energies are the resonances $\triangle$(1232), N*(1440) and the pions. The reaction channels are given as follows:
\begin{eqnarray}
&& NN \leftrightarrow N\triangle, \quad  NN \leftrightarrow NN^{\ast}, \quad  NN
\leftrightarrow \triangle\triangle,  \nonumber \\
&& \Delta \leftrightarrow N\pi,  N^{\ast} \leftrightarrow N\pi,  NN \rightarrow NN\pi (s-state).
\end{eqnarray}
The cross sections of each channel to produce resonances are
parameterized by fitting the data calculated with the one-boson
exchange model \cite{Hu94}. In the 1 A GeV region, there are mostly
$\Delta$ resonances which disintegrate into a $\pi$ and a nucleon,
however, the $N^{\ast}$ yet gives considerable contribution to the
high energetic pion yield. The energy and momentum dependent decay
width is used in the calculation \cite{Fe09}. The strangeness is created by inelastic hadron-hadron collisions. We include the
channels as follows:
\begin{eqnarray}
&& BB \rightarrow BYK,  BB \rightarrow BBK\overline{K},  B\pi \rightarrow YK,  B\pi \rightarrow NK\overline{K}, \nonumber \\
&& Y\pi \rightarrow N\overline{K}, \quad  N\overline{K} \rightarrow Y\pi, \quad YN \rightarrow \overline{K}NN.
\end{eqnarray}
Here the B strands for (N, $\triangle$) and Y($\Lambda$, $\Sigma$), K(K$^{0}$, K$^{+}$) and $\overline{K}$($\overline{K^{0}}$, K$^{-}$). The parameterized cross sections of each isospin channels $BB \rightarrow BYK$ \cite{Ts99} are used in the calculation. We take the parametrizations of the channels $B\pi \rightarrow YK$ \cite{Ts94} besides the $N\pi \rightarrow \Lambda K$ reaction \cite{Cu84}. The results are close to the experimental data at near threshold energies. The cross section of antikaon production in inelastic hadron-hadron collisions is taken as the same form of the parametrization used in hadron string dynamics (HSD) calculations \cite{Ca97}. Furthermore, the elastic channels are considered through the channels $KB \rightarrow KB$ and $\overline{K}B \rightarrow \overline{K}B$ and we use the parametrizations in Ref. \cite{Cu90}.

\begin{figure*}
\includegraphics{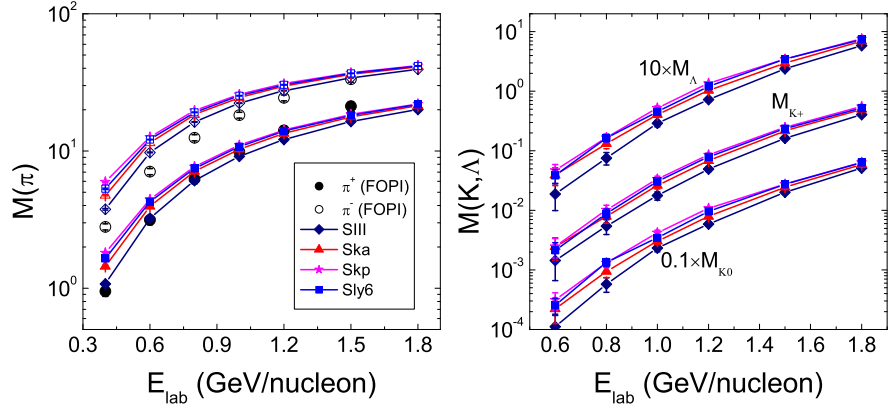}
\caption{\label{fig:wide} (Color online) Excitation functions of charged pions and strangeness production in central $^{197}$Au+$^{197}$Au collisions at different Skyrme forces.}
\end{figure*}

\begin{figure*}
\includegraphics{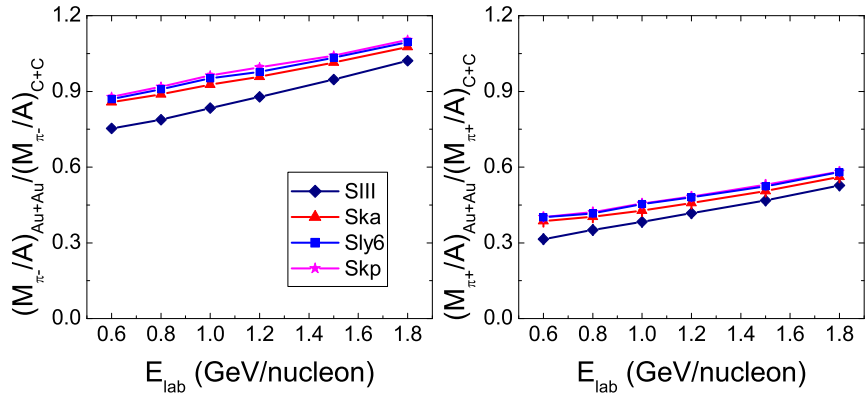}
\caption{\label{fig:wide} (Color online) Comparison of the calculated excitation functions of the ratios of charged pion multiplicities obtained in $^{197}$Au+$^{197}$Au over $^{12}$C+$^{12}$C reactions for head-on collisions at different Skyrme forces.}
\end{figure*}

The dynamics of pion and strangeness production in heavy-ion collisions in the region of 1A GeV energies is investigated systematically by using the LQMD model. Shown in Fig. 1 is a comparison of charged pions and strange particles in the $^{197}$Au+$^{197}$Au reactions and compared with the available data measured by FOPI collaboration \cite{Re10}. The charged pions are overpredicted at near threshold energies, in particular for the $\pi^{-}$ production, which results from the different stiffness of the symmetry energy \cite{Fe10a,Fe10b}. A similar trend is clear in the strangeness production at different Skyrme forces. Overall, the parameter SIII leads to lower yields of strange particles. In order to reduce some uncertainties, such as cross sections, in-medium decay width of resonances etc., we calculated the double ratios of charged pions per mass numbers of reaction partners produced in $^{197}$Au+$^{197}$Au over $^{12}$C+$^{12}$C collisions as shown in Fig. 2. A slight increase of the value with incident energy is obvious for each parameters, which is consistent with the experimental observations \cite{St01}. One can see that the influence of the equation of state on the excitation functions is slightly. Therefore, it is difficulty to extract the high-density information of nuclear equation of state from pion emissions.

\begin{figure}
\includegraphics[width=8 cm]{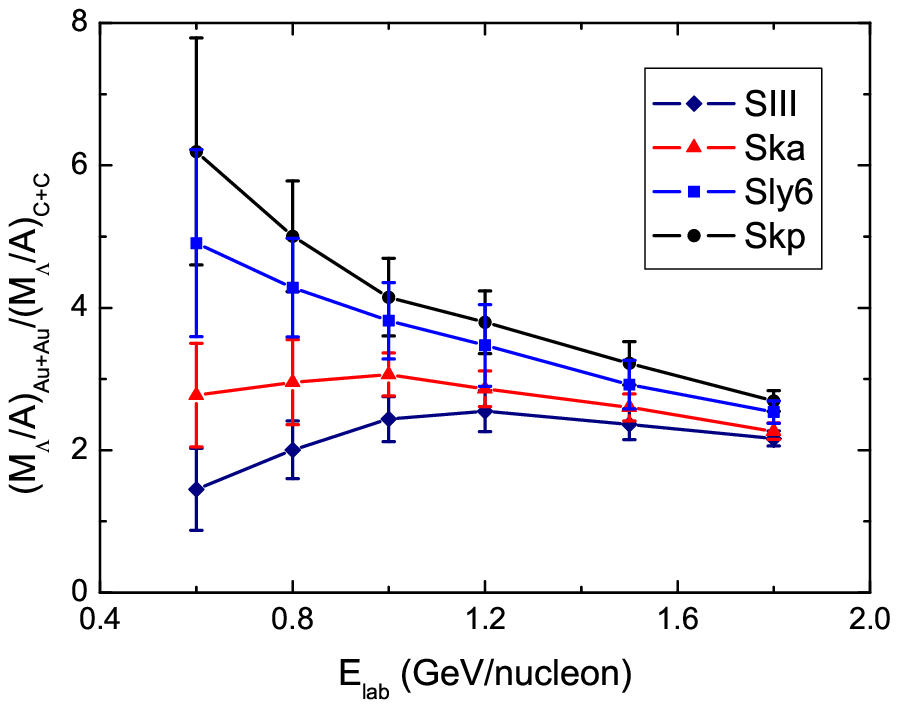}
\caption{\label{fig:epsart} (Color online) The same as in Fig. 2, but for the case of $\Lambda$ production.}
\end{figure}

Neutral strange particles can not be influenced by the Coulomb interaction of surrounding baryons. Production of $\Lambda$ is of great interest as a probe of high-density information of nuclear matter. We calculated the excitation functions of double ratios of $\Lambda$ emissions in heavy-ion collisions as shown in Fig. 3. Different distributions are pronounced for each Skyrme forces. The parameters Sly6 and SkP give a monotonically decreasing trend with increasing incident energy. However, the forces SIII and Ska reduce the values and even slightly drop at low energy, which correspond to a hard equation of state of symmetric nuclear matter. Experimental measurements are helpful for constraining the stiffness of nuclear EoS. The increase of the ratios with decreasing incident energy is caused by the fact that the number of collisions where the involved particles are encountered prior to the production of a kaon does not reach a sharp region for the system C+C at far low energies, which is related to higher compressible nuclear matter.

\begin{figure*}
\includegraphics{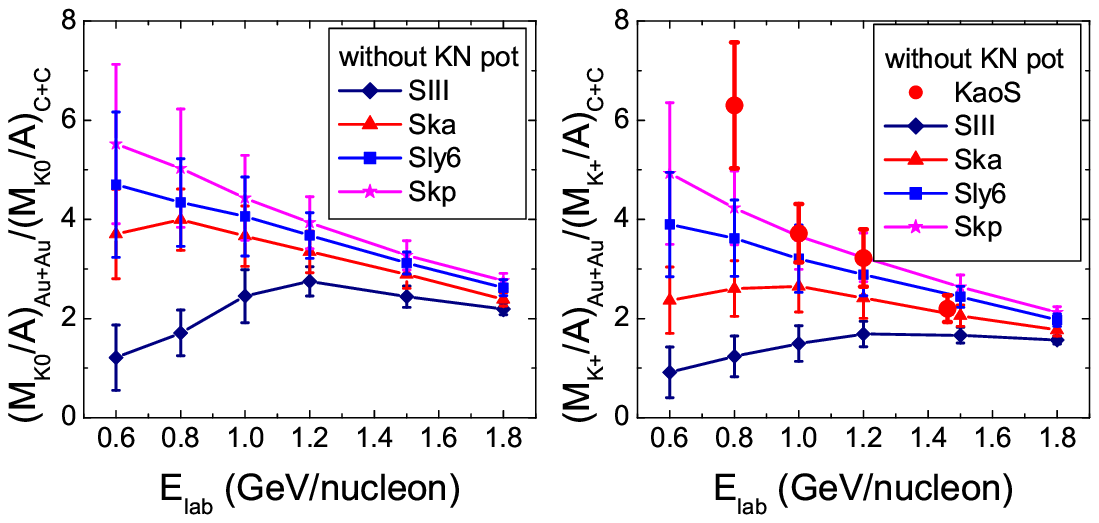}
\caption{\label{fig:wide} (Color online) Excitation functions of kaon multiplicities calculated by the LQMD model without $KN$ potential in central $^{197}$Au+$^{197}$Au over $^{12}$C+$^{12}$C collisions and compared with the available data from KaoS collaboration for $K^{+}$ production.}
\end{figure*}

\begin{figure*}
\includegraphics{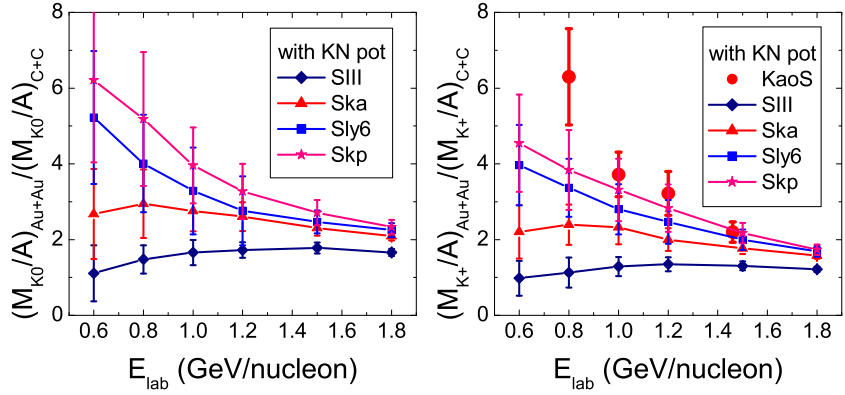}
\caption{\label{fig:wide} (Color online) The same as in Fig. 4, but for calculations with $KN$ potential.}
\end{figure*}

\begin{figure}
\includegraphics[width=8 cm]{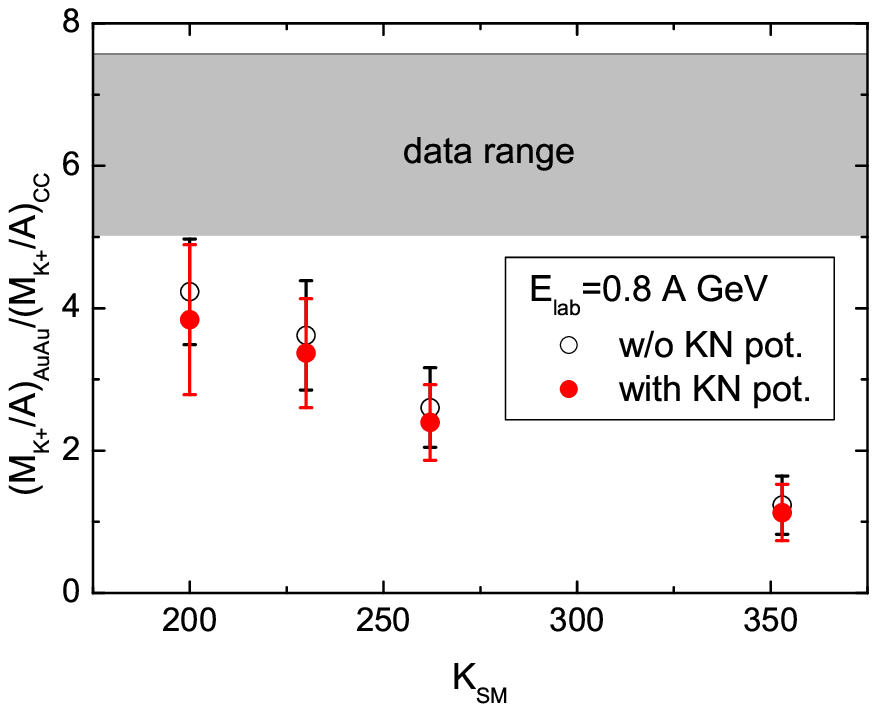} \includegraphics[width=8 cm]{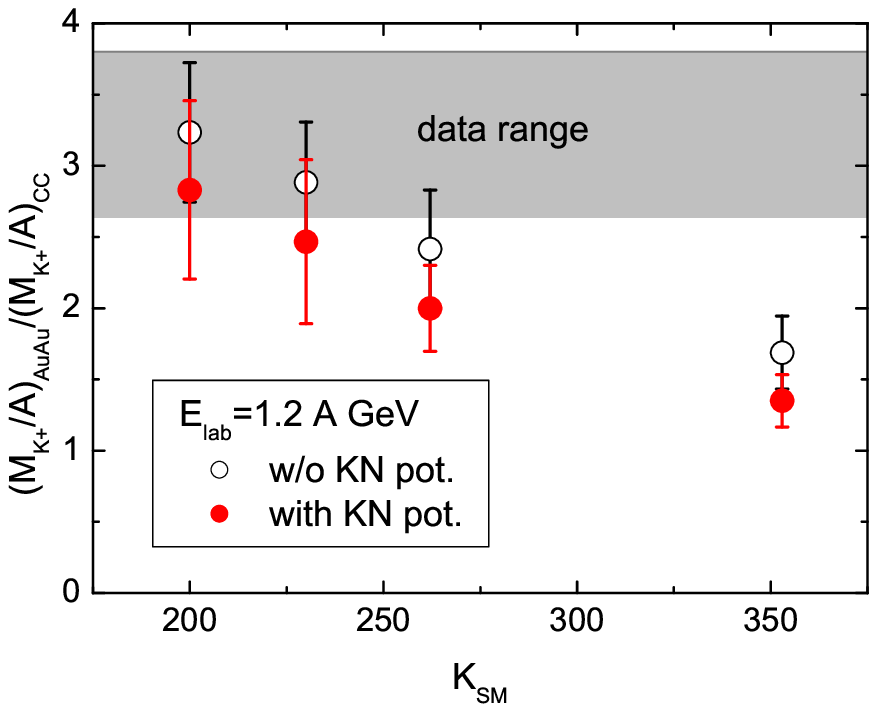}
\caption{\label{fig:epsart} (Color online) Double ratios calculated within the LQMD model at 0.8A GeV (up panel) and 1.2A GeV (down panel) as a function of the incompressibility coefficient $K_{SM}$ and compared with the available data for the K$^{+}$ production.}
\end{figure}

To extract more information of the high-density EoS, in Fig. 4 we show the kaon production in the $^{197}$Au+ $^{197}$Au and $^{12}$C+$^{12}$C reactions for head-on collisions, normalized by the corresponding mass numbers and compared with the KaoS data \cite{St01}. A similar distribution with the $\Lambda$ emission in Fig. 3 is observed, which results from the same channels for the kaon and $\Lambda$ production, such as $BB \rightarrow B\Lambda K$ and $B\pi \rightarrow \Lambda K$. A larger high-density region ($>\rho_{0}$) is formed in $^{197}$Au+ $^{197}$Au collisions and the compression depends on the nuclear equation of state. Whereas the compression in $^{12}$C+$^{12}$C reaction is small and not sensitive to the stiffness of the EoS. The comparison to the available data further supports a soft equation of state in the high-density region, in particular the parameter SkP, where the increase of the double ratios with decreasing incident energy is more pronounced when going far below threshold energies. Inclusion of the in-medium kaon-nucleon (KN) potential as shown in Fig. 5 almost does not change the results. The conclusions are consistent with the calculations by Fuchs \emph{et al.} within the QMD model \cite{Fu01} and by Hartnack \emph{et al.} within the Isospin-QMD (IQMD) model \cite{Ha06}. Figure 6 is the incompressibility dependence of the double ratios at incident energies 0.8A GeV and 1.2A GeV. The KN potential slightly reduces the ratios because of the decrease of the secondary collision rate for the production of kaon. Furthermore, the soft equation of state is close to the experimental data.

In summary, the production of pions and strange particles in heavy-ion collisions in the region of 1A GeV energies as probes of the nuclear EoS is investigated within the framework of the LQMD model. The charged pions of the available experimental data can be reproduced well within the selected Skyrme parameters SkP, SLy6, Ska and SIII. The excitation functions of the double ratios of charged pions obtained in the $^{197}$Au+$^{197}$Au over $^{12}$C+$^{12}$C collisions weakly depend on the nuclear equation of state. Whereas, the ratios of strange particle production are sensitive to the potential parameters of nucleon-nucleon interaction. The comparison to the KaoS data favors a soft equation of state. This conclusion can not be modified by the input parameters, such as the KN potential, the lifetime of the $\triangle$ in the nuclear matter, cross sections in production of strangeness.

This work was supported by the National Natural Science Foundation
of China under Grant 10805061; the Special Foundation of the
President Fund; the West Doctoral Project of Chinese Academy of
Sciences; and the Major State Basic Research Development Program
under Grant 2007CB815000.

\end{document}